\documentclass[physics,article,accept,pdftex,moreauthors]{Definitions/mdpi} 
\graphicspath{{Definitions/}}
\usepackage{graphicx}

 \newcommand\la{\langle}
 \newcommand\ra{\rangle}
 \newcommand\beq{\begin{equation}}
 
 \newcommand\eeq{\end{equation}}
 \newcommand\beqn{\begin{eqnarray}}
 \newcommand\eeqn{\end{eqnarray}}
 \newcommand\GeV{{\rm GeV}}

\def\Re{\,{\rm Re}\,}
\def\Im{\,{\rm Im}\,}

\def\fm{\,\mbox{fm}}
\def\GeV{\,\mbox{GeV}}

\def\Pom{{\rm I\!P}}

\def\lsim{\mathrel{\rlap{\lower4pt\hbox{\hskip1pt$\sim$}}
    \raise1pt\hbox{$<$}}}         
\def\gsim{\mathrel{\rlap{\lower4pt\hbox{\hskip1pt$\sim$}}
    \raise1pt\hbox{$>$}}}         

\def\la{\langle}
\def\ra{\rangle}

\firstpage{1} 
\pubvolume{1}
\issuenum{1}
\articlenumber{0}
\pubyear{2023}
\copyrightyear{2023}
\datereceived{11 September 2023} 
\daterevised{12 October 2023} 
\dateaccepted{ } 
\datepublished{ } 
\hreflink{https://doi.org/} 

\Title {Single-Spin Asymmetry of Neutrons in Polarized ${pA}$ 
Collisions}

\TitleCitation {Single-Spin Asymmetry of Neutrons in Polarized $pA$ Collisions}
  
\Author{{Boris 
Z.} 
 Kopeliovich *, {Irina 
K.} Potashnikova and Iv\'an Schmidt }

\address[1]{Departamento de F\'{\i}sica,
Universidad T\'ecnica Federico Santa Mar\'{\i}a,
Avenida Espa\~na 1680, 
239-0123 {Valpara\'iso,} 
 Chile; {irina.potashnikova@usm.cl (I.K.P.); ivan.schmidt@usm.cl (I.S.)} 
}

\AuthorNames{Firstname Lastname, Firstname Lastname and Firstname Lastname}

\AuthorCitation{{Kopeliovich,} 
 B.Z.; Potashnikova, I.K.; Schmidt, I.}

\corres{\hangafter=1 \hangindent=1.05em \hspace{-0.82em} Correspondence: boris.kopeliovich@usm.cl}

\abstract{
Absorptive corrections, which are known to suppress proton-neutron transitions with a large fractional momentum
$z\to1$ in $pp$ collisions, become dramatically strong on a nuclear target, and they push the partial cross sections of leading neutron production to the very periphery of the nucleus.
The mechanism of the pion $\pi$ and axial vector meson $a_1$
interference, 
which successfully explains the observed single-spin asymmetry in a polarized $pp\to nX$, is extended to the collisions of polarized protons with nuclei. When corrected for nuclear effects, it explains the observed single-spin azimuthal asymmetry of neutrons that is produced in inelastic events, which is where the nucleus violently breaks up. This single-spin asymmetry is found to be negative and 
nearly {atomic mass number}   
$A$-independent.}


\begin{document}

\section{Proton-Neutron Transitions in the Vicinity of the Pion Pole}

The pion is known to have a large coupling to nucleons; therefore,
pion exchange is important in processes with isospin one in the
cross channel (e.g., $p+n\to n+p$). However, the Regge pion
trajectory has a low intercept of $\alpha_\pi(0)\approx 0$, and this is
why it ceases to be important at high energies in binary reactions.
Meanwhile, other mesons, such as $\rho,\ a_2$, etc., take over.
More specifically, in the inclusive production of neutrons, the whole rapidity interval is divided between the rapidity gap and the exited multiparticle state $X$ with invariant mass. The latter might be sufficiently large enough to reduce the rapidity gap, i.e., what allows pion exchange.

\section{A Triple-Regge Description of Neutron Production }

\begin{figure}[H]
  \scalebox{0.3}{\includegraphics{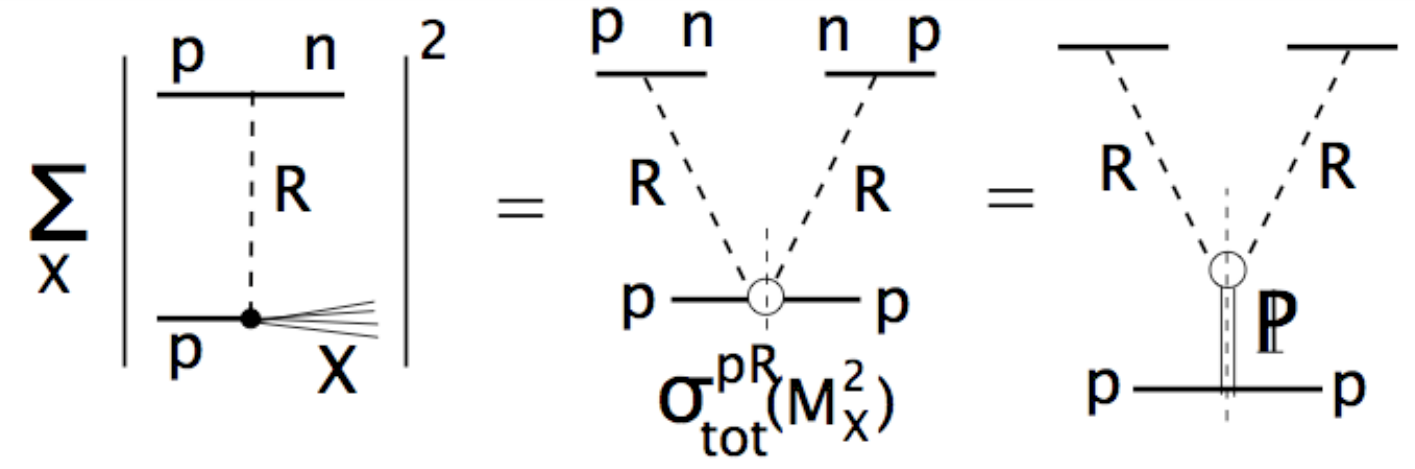}}
\caption{\label{fig:3R} Graphical relation between 
the cross section of neutron production and the total Reggeon-proton cross section, 
$\sigma^{pR}_{\rm{tot}}$, which—when with a large invariant mass,
$M_X^2$, of the final states—is dominated by the Pomeron, $\mathbb P$}.
 \end{figure}

The process $p\!\uparrow+p(A)\to n+X$
that has a large, fractional light-cone momentum $z=p^+_n/p^+_p$
of neutrons produced that is in the proton beam direction {{(rather, in the target direction, i.e., the asymmetry is zero according to the Abarbanel-Gross theorem}
~\cite{gross}{)}} , is known to
be related to the iso-vector Reggeons ($\pi,\ \rho,\ a_2,\ a_1$, etc.) 
\cite{kpss}. Here, the upward arrow denotes that the proton is transversely polarized, and the '$+$'  
sign {denotes the plus-component of the particle ligh-cone momentum.} 
As illustrated in Figure~\ref{fig:3R}, the amplitude, squared, and summed over values of the final states $X$ (at a fixed invariant mass, $M_X$) are expressed via the Reggeon-proton total cross section at the 
{centre-of-mass 
(c.m.)} 
energy $M_X$.

At high energies of colliders (RHIC (Relativistic Heavy Ion Collider), LHC (Large Hadron Collider)
and others),
 the $M_X^2=s(1-z)$, where $s$ denotes the collision c.m. energy squared, 
is so large (except in the inaccessibly small $1-z$) that the cross section 
$\sigma^{pR}_{\rm{tot}}(M_X^2)$ 
is dominated by the Pomeron exchange, as is illustrated in Figure~\ref{fig:3R}.
The couplings of the iso-vector Reggeons, specifically those with a natural parity ($\rho$, $a_2$) with the proton, are known to be predominantly spin-flip~\cite{kane}; as such, they can be neglected, and this is the case because we are interested here in the small transverse 
momenta, $p_T\to0$, 
of the neutrons. 
Only unnatural parity Reggeons ($\pi$, $a_1$) have large spin non-flip couplings that contribute in the forward direction.

 %
 \section{Cross Section of Forward Neutron Production}
\subsection{Pion Pole}

Pions are known to have a large coupling with nucleons; thus, the
pion exchange is important in processes with isospin flip, like in $p\to n$. 
Measurements with polarized proton beams supply more detailed information about the interaction dynamic. The cross section of neutron production in the vicinity of the pion pole was measured at the 
 ISR (the Interacting Storage Rings) 
in $pp$ collisions~\cite{isr} and in 
deep inelastic scattering (DIS) at HERA (Hadron-Electron Ring Accelerator)~\cite{k2p}.
The cross section has also been calculated within various 
theoretical schemes~\cite{bishari, boreskov1, boreskov2,ap,strong1,strong2,ryskin}.

The pion term in the cross section of neutron production reads as follows~\cite{kpss}:
\beq
\left.z\,\frac{d\sigma^B(pp\to nX)}{dz\,dq_T^2}\right|_{\pi}=
f_{\pi/p}(z,q_T,q_L)\,
\sigma^{\pi^+ p}_{\rm{tot}}(s').
\label{100}
 \eeq
 
Here,
 {$z$}  is related to 
 {$M_X$} 
 and 
$z\approx
1-M_X^2/s$ for $1-z\ll1$, where $\sqrt{s}$ is the c.m. energy of $pp$ 
collision.
Furthermore, $s'=M_X^2$ and the superscript $B$ means  
that this is a Born approximation; thus, the absorptive corrections will be ignored and considered later on. 

$f_{\pi/p}(z,q_T,q_L)$ is the pion flux in the proton, and it has the form
 \beqn
f_{\pi/p}(z,q_T,q_L)&=&
|t|\,G_{\pi^+pn}^2(t)\left|\eta_\pi(t)\right|^2
\left(\frac{\alpha_\pi^\prime}{8}\right)^2 
\nonumber\\ &\times&
(1-z)^{1-2\alpha_\pi(t)}.
\label{120}
 \eeqn
 
Here, $q_T$ is the neutron transverse momentum,
$
q_L=m_N(1-z)/\sqrt{z}
$, 
with $m_N$ denoting the nucleon mass,
$
-t=q_L^2+q_T^2/z
$, 
{$\alpha_\pi^\prime$ is the slope of the pion Regge trajectory,} 
and $G_{\pi^+pn}(t)$ is the effective 
 {$\pi N$} 
vertex function~\cite{k2p,kpss}.
$G_{\pi^+pn}(t)=g_{\pi^+pn}\exp(R_{\pi NN}^2t)$
includes the pion-nucleon coupling
and the form factor which incorporates the t-dependence
of the coupling and of the $\pi N$ inelastic amplitude.
We take the values of the parameters used in Ref. \cite{k2p}
$g_{\pi^+pn}^2/8\pi=13.85$ and $R_{\pi NN}^2=0.3\GeV^{-2}$.
 Notice that
the choice of  $R_{\pi NN}$ does not bring much uncertainty, since
we focus here at data for forward production, $q_T=0$, so
$t$ is quite small.
The signature factor is 
$\eta_\pi(t)=i-
\cot  
\left[\pi\alpha_\pi(t)/2\right]$.

The amplitude of the process includes both non-flip and spin-flip terms as follows~\cite{kpss,kpss-spin}:
\beq
A^B_{p\to n}(\vec q,z)=
\bar\xi_n\left[\sigma_3\, q_L+
\frac{1}{\sqrt{z}}\,
\vec\sigma\cdot\vec q_T\right]\xi_p\,
\phi^B(q_T,z),
\label{900}
\eeq
where $\vec\sigma$ are the Pauli matrices, and $\xi_{p,n}$ are the proton or
neutron spinors.

In the region of a small $1-z\ll1$, the pseudo-scalar amplitude $\phi^B(q_T,z)$ has 
the 
triple-Regge form (see Figure~\ref{fig:3R}) as follows:
 \beqn
\phi^B(q_T,z)&=&\frac{\alpha_\pi^\prime}{8}\,
G_{\pi^+pn}(t)\,\eta_\pi(t)\,
(1-z)^{-\alpha_\pi(t)}\nonumber\\ &\times&
A_{\pi p\to X}(M_X^2).
\label{122}
 \eeqn

The procedure of inclusion of the absorptive corrections on the amplitude level was developed in 
{Refs.}~\cite{kpss,kpps}. First, the amplitude (\ref{900}) is Fourier transformed to impact parameter representation, which is where the 
absorptive 
effect is just a multiplicative suppression factor. Then, the absorption-corrected amplitude is Fourier transformed back to a momentum representation. The effects of absorption turn out to be quite strong, i.e., it 
roughly reduces the neutron production cross section two-fold, and it differently affects the non-flip and spin-flip terms in the amplitude.Details of the calculations can be found in~\cite{kpss,kpps,pi-pi}.

Notice that the contributions of other iso-vector Reggeons, such as $\rho$, $a_2$, and the effective $\tilde a_1$ 
were calculated in Ref.~\cite{kpps}. For the kinematics of data under discussion
~\cite{prl}, $q_T^2\sim 0.01\GeV^2$, $\la z\ra=0.75$,$\rho$, and $a_2$ were found to be vanishingly small (see Figure~8 in Ref.~\cite{kpps}). The contribution of $\tilde a_1$ was small as well, but it did not contribute
to the single-spin asymmetry.

There was also a contribution of neutrons to the production and decay of $\Delta$ resonance as well as to other resonances. This phenomenon was carefully evaluated in 
Ref.~\cite{k2p}, whereby the results of a phase-shift analysis of 
$\pi N$  
was employed by scattering at low energies. This correction was also found to be small in
in {Ref.}~\cite{k2p}.
Therefore, the dominance of the pion is well justified.
\subsection{Absorptive Corrections}

\begin{figure}[H]
  {\includegraphics[width=6cm]{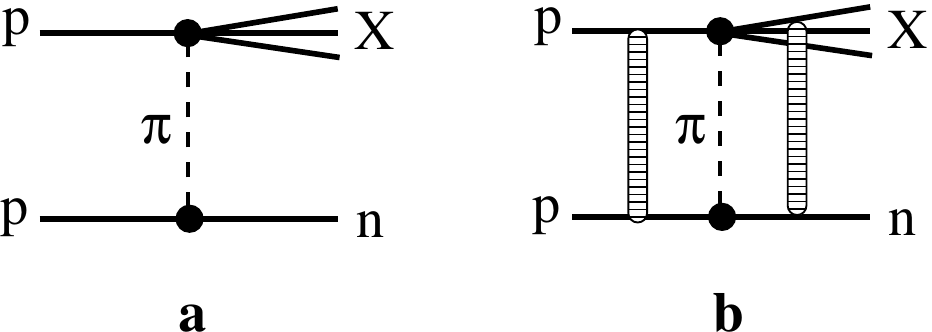}}
\caption{\label{abs} Graphical 
 illustration of the absorptive effects in inclusive neutron production in 
proton-proton {collisions:} 
({\bf a}) 
the pion exchange 
{(
dashed line)} in the Born approximation; 
({\bf b}) absorptive corrections due to initial and final state interactions shown by a 
 {ladder}-type exchanges.}
 \end{figure}

Like any process with LRGs,  forward neutron production is subject to the large absorptive corrections that come from initial/final state interactions, as illustrated in Figure~\ref{abs}.

The evaluation of the corrections is quite complicated in momentum
representation, i.e., where they require multi-loop integrations. However,
if these corrections do not correlate with the amplitude of the process
$\pi^+p\to X$, then they factorize in impact parameters
and become much simpler. Therefore, first of all, one should Fourier transform the amplitude Equation~(\ref{900}) to impact the parameter space.

The partial Born amplitude at impact parameter $\vec b$, which
corresponds to {Equation} 
(\ref{100}) for the differential cross section, has the form~\cite{kpss}

 \beq
f^B_{p\to n}(\vec b,z)=\frac{1}{\sqrt{z}}\,
\bar\xi_n\left[\sigma_3\,\tilde q_L\,\theta^B_0(b,z)-
i\,\frac{\vec\sigma\cdot\vec b}{b}\,
\theta^B_s(b,z)\right]\xi_p\,,
\label{150}
 \eeq
where 
\beqn
\theta^B_0(b,z) &=& \int d^2q\,e^{i\vec b\vec q}\,
\phi^B(q_T,z)
\nonumber\\ &=&
\frac{N(z)}{1-\beta^2\epsilon^2}\,
\left[K_0(\epsilon b)-K_0(b/\beta)\right]\, , 
\label{154}
 \eeqn

 \beqn
\theta^B_s(b,z) &=& {1\over b}
\int d^2q\,e^{i\vec b\vec q}\,
(\vec b\cdot\vec q)\,\phi^B(q_T,z)
\nonumber\\ &=&
\frac{N(z)}{1-\beta^2\epsilon^2}\,
\left[\epsilon\,K_1(\epsilon b)-\frac{1}{\beta}\,K_1(b/\beta)\right]\,.
\label{164}
 \eeqn

 {Here, $K_{0,1}$ are the modified Bessel functions of the zero and first orders,  
respectively, and} 
 \beqn
N(z) &=&\frac{1}{2}\,g_{\pi^+pn}\,
z(1-z)^{\alpha^\prime_\pi(m_\pi^2+\tilde q_L^2/z)}
e^{-R_{\pi NN}^2 \tilde q_L^2/z}
\nonumber\\&\times&
A_{\pi p\to X}(M_X^2)\,
\nonumber\\
\epsilon^2&=&\tilde q_L^2+zm_\pi^2\,,
\nonumber\\
\beta^2&=&R_{\pi NN}^2-\alpha_\pi^\prime\,\frac{\ln(1-z)}{z}
\, .
\label{166}
 \eeqn

To simplify the calculations we replaced here, the Gaussian form factor
$\exp(-\beta^2q_T^2)$ is replaced with the monopole form $1/(1+\beta^2q_T^2)$,
which is a suitable approximation of the small values of $q_T$ that we are interested in.

 The corrected cross section is then calculated as a convolution of the cross section with the 
survival probability amplitude 
factor:   
 \beq
f_{p\to n}(b,z)=f^B_{p\to n}(b,z)\,S(b,z)\,, 
\label{195}
 \eeq
where $S(b,z)$ is the survival amplitude. 

The process under consideration, i.e., that the large $z\to1$ is associated with
the creation of a large rapidity gap (LRG) $\Delta y=|\ln(1-z)|$, which is where no
particle is produced. Absorptive corrections may be also interpreted as a
suppression that is related to the survival probability of LRG. 

The process under consideration contains the amplitude of the pion-proton inelastic
collision $\pi+p\to X$. The latter is typically described as a color exchange that leads to the creation of two color octet states with a large rapidity
interval $\sim \ln(M_X^2/s_0)$ ($s_0=1\GeV^2$). 
 Perturbatively, the interaction is mediated by gluonic exchanges.
Nonperturbatively, e.g., in the string model, the hadron collision looks like there is a crossing
and flip of the strings.

One may think that this is the produced color
octet-octet state, which experiences the final state interactions with
the recoil neutron. On the other hand, at high energies, multiple 
interactions become coherent, and both initial and final state 
interactions must be included. This leads to a specific space-time 
development of the process at high
energies; namely, the incoming proton fluctuates into a five-quark {state,}  
$|\{3q\}_8\{\bar qq\}_8\ra$, which is long, in advance, in its interaction with the target via pion exchange.

We evaluate 
$S(b,z)$ with the color dipole model.
One can present the survival amplitude of a five-quark state with an accuracy of $1/N_c^2$ as
 \beq
S^{(5q)}(b) = S^{(3q)}(b)\,S^{(q\bar q)}(b)
=\left[1-\Im\Gamma^{(3q)p}(b)\right]
\left[1-\Im\Gamma^{(\bar qq)p}(b)\right]\,.
\label{290}
 \eeq
   
The elastic amplitude $\Gamma^{(\bar33)p}(b)$ of a color $\{\bar 33\}$    
dipole interacting with a proton is related to the partial elastic amplitude~\cite{kpss},
 \beq
\Im\Gamma^{(\bar 33)p}(b,z)=
\int d^2r W_{\bar 33}(r,M_X^2)\,
\Im f^{\bar33}_{\rm{el}}(\vec b,\vec r,s,\alpha),
\label{295}
 \eeq
 where $\alpha$ is the fractional light-cone momentum carried by the $3$ or $\bar
3$ triplets, 
 $r$ is the dipole transverse size and $W_{\bar 33}(r,M_X^2)$ is the
dipole size distribution function~\cite{kpss}.

\section{Azimuthal Asymmetry of Neutrons in \boldmath{$pp$} Collisions}

\subsection{Born Approximation}

Both terms in the {amplitude 
(\ref{900})} 
have the same phase factor, $\eta_\pi(t)$.
Therefore, in spite of the presence of both spin-flip and non-flip amplitudes, no single-spin asymmetry associated with pion exchange is possible in the Born approximation. Even the inclusion of absorptive corrections leave the spin effects miserably small~\cite{kpss-trieste,kpss-spin} compared to \mbox{the data~\cite{phenix-pp1,phenix-pp2}. }

A plausible candidate for generating a sizable spin asymmetry at high energies is the $a_1$ meson exchange since $a_1$ can be produced by pions diffractively. However, this axial-vector resonance 
is hardly visible in the diffractive channels $\pi+p\to 3\pi+p$, which are dominated by $\pi\rho$
in the $1^+S$ wave. The $\pi\rho$ invariant mass distribution forms a pronounced narrow peak at the invariant mass
$M_{\pi\rho}\approx m_{a_1}$ (due to the Deck effect~\cite{deck}). Although, in the dispersion relation for the amplitude, this channel corresponds to a cut, it can be replaced with 
good
accuracy by an effective pole $\tilde a_1$~\cite{belkov,kpss-spin}.
In the crossed channel, $\pi\rho$ exchange corresponds to a Regge cut with the known intercept and slope of the Regge trajectory~\cite{kpss-spin}.

The expression for the single-spin asymmetry arises from the 
$\pi\,\tilde a_1$  
interference, which has the form~\cite{kpss-spin}
\beqn
&&A_N^{(\pi\, \tilde a_1)}(q_T,z) =
q_T\,\frac{4m_N\,q_L}{|t|^{3/2}}\,
(1-z)^{\alpha_\pi(t)-\alpha_{\tilde a_1}(t)}
\label{920}
\\ &\times&
\frac{\Im\,\eta_\pi^*(t)\,\eta_{\tilde a_1}(t)}
{\left|\eta_\pi(t)\right|^2}\,
\left(\frac{d\sigma_{\pi p\to \tilde a_1p}(M_X^2)/dt|_{t=0}}
{d\sigma_{\pi p\to\pi p}(M_X^2)/dt|_{t=0}}\right)^{1/2}
\frac{g_{\tilde a_1^+pn}}{g_{\pi^+pn}}.
\nonumber
\eeqn

The trajectory of the $\pi\rho$ Regge cut and the phase factor $\eta_{\tilde a_1}(t)$ are known from Regge phenomenology.
The ${\tilde a_1}NN$ coupling was evaluated in Ref.~\cite{kpss-spin}, and it is based on 
the partial conservation of axial current
(PCAC)  and the 
second Weinberg sum rule, where the spectral functions of the vector and axial
currents are represented by the $\rho$ and the effective ${\tilde a_1}$ poles, respectively.
This leads to the following relations between the couplings:
\beq
\frac{g_{\tilde a_1 NN}}{g_{\pi NN}}=
\frac{m_{\tilde a_1}^2\,f_\pi}{2m_N\,f_\rho}\approx {1\over2},
\label{340}
\eeq
where $f_\pi=0.93m_\pi$ is the pion decay coupling,  
$f_\rho=\sqrt{2}m_\rho^2/\gamma_\rho$ 
and $\gamma_\rho$ is the universal coupling: $\gamma_\rho^2/4\pi=2.4$.

\subsection{Absorption-Corrected Spin Amplitudes}

The absorption corrections to the spin amplitudes in Equation~(\ref{150}) with respect to the impact parameter representation are calculated in analogy to Equation (\ref{195}):
\beq
\tilde \theta_{0,s}(b,z)= \theta^B_{0,s}(b,z)\ S_{0,s}(b,z),
\label{sbs-spin}
\eeq
where the absorption factors $S_{0,s}(b,z)$ are different for the spin amplitudes, as is explained and evaluated in Ref.~\cite{kpss}.

By performing an inverse Fourier transformation back to momentum representation, 
one obtains  
the absorption-corrected spin amplitudes
\beq
A_{p\to n}(\vec q,z)=
\bar\xi_n\left[\sigma_3 q_L\,\phi_0(q_T,z)-
i\vec\sigma\cdot\vec q_T\frac{\phi_s(q_T,z)}{\sqrt{z}}\right]\xi_p,
\label{522}
 \eeq
 where the amplitudes $\phi_{0,s}$ are given by the inverse Fourier transformation back to momentum representation:
   \beqn
\Re\phi_0(q_T,z)&=&\frac{N(z)}{2\pi(1-\beta^2\epsilon^2)}
\int\limits_0^\infty db\,b\,J_0(bq_T)
\left[K_0(\epsilon b)-K_0\left({b\over\beta}\right)\right]
S(b,z)\,, 
\nonumber\\
\Im\phi_0(q_T,z)&=&\frac{\alpha_\pi^\prime N(z)}{4z\beta^2}
\int\limits_0^\infty db\,b\,J_0(bq_T)\,
K_0\left({b\over\beta}\right)\,
S(b,z)\,, 
\label{550}
 \\
 q_T\Re\phi_s(q_T,z)&=&\frac{N(z)}{2\pi(1-\beta^2\epsilon^2)}  
\int\limits_0^\infty db\,b\,J_1(bq_T)
\left[\epsilon\, K_1(\epsilon b)-
{1\over\beta}\,K_1\left({b\over\beta}\right)\right]   
S(b,z)\, , 
\nonumber\\
q_T\Im\phi_0(q_T,z)&=&\frac{\alpha_\pi^\prime N(z)}{4z\beta^3}
\int\limits_0^\infty db\,b\,J_1(bq_T)\,
K_1\left({b\over\beta}\right)\,   
S(b,z)\,.
\label{570}
 \eeqn 
 {Here $J_{0,1}$ are the Bessel functions of the zero and first orders, respectively.} 

Eventually, the single-spin asymmetry obtains the form
 \beqn
A_N(q_T,z)&=&\frac{2q_T q_L\sqrt{z}
\sum_X
\left|\phi_0(q_T,z)\right|
\left|\phi_s(q_T,z)\right|}
{z q_L^2 \sum_X\left|\phi_0(q_T,z)\right|^2 +
q_T^2 \sum_X\left|\phi_s(q_T,z)\right|^2}
\nonumber\\ &\times&
\sin(\delta_s-\delta_0)\,,
\label{622}
 \eeqn
 where
 \beq
\tan\delta_{0,s}=\frac{\Im\phi_{0,s}(q_T,z)}
{\Re\phi_{0,s}(q_T,z)}\,.
\label{640}
 \eeq

\begin{figure}[H]
 {\includegraphics[height=6cm]{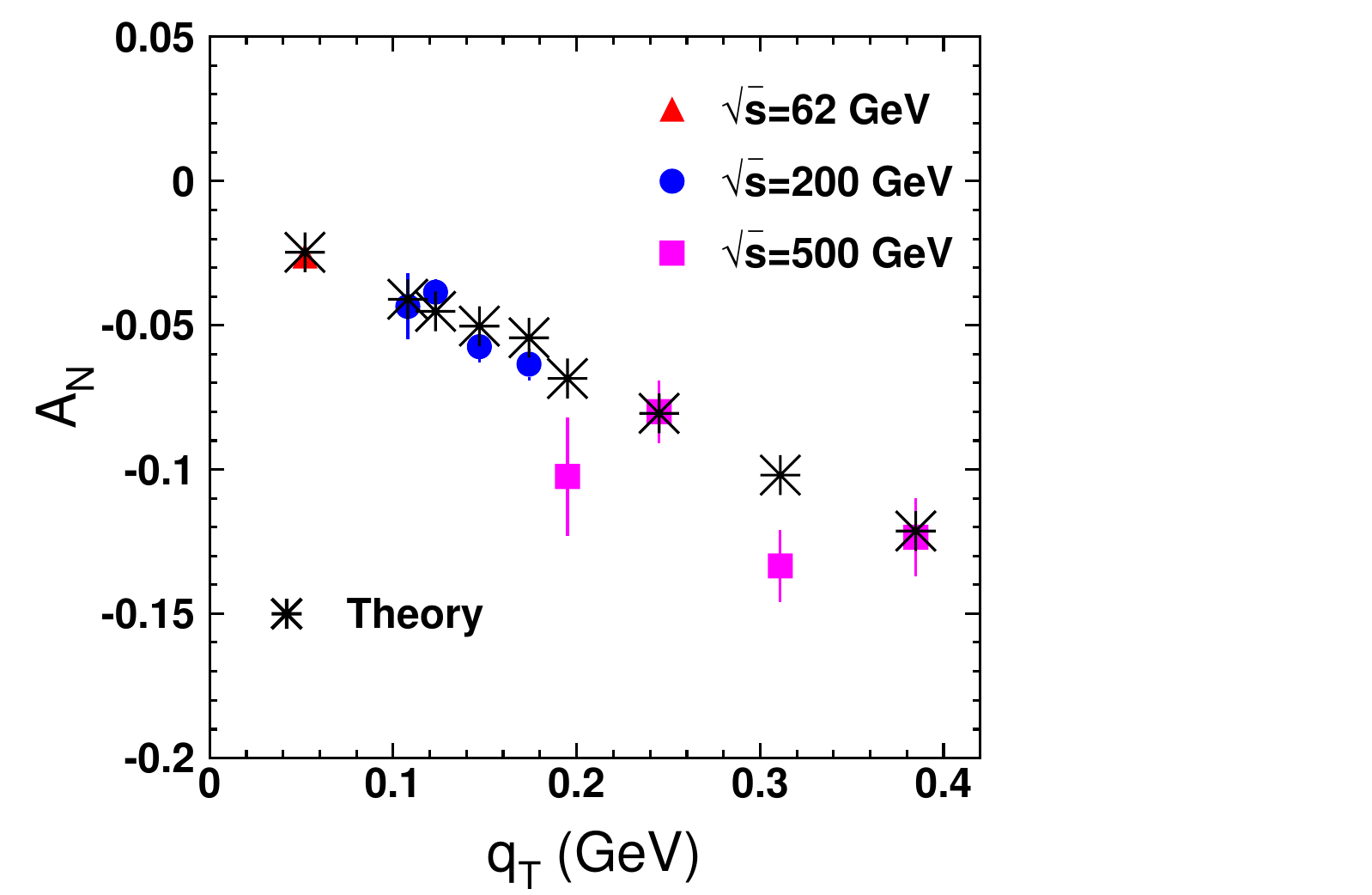}}
\caption{\label{fig:AN-pi-a}  {The single-spin} 
 {asymmetry, $A_N$,}  
in the polarized $pp\to nX$ process 
at the centre-of-mass energy 
$\sqrt{s}=200\GeV$ {versus} 
neutron transverse momentum $q_T$. The data points from
Refs.~\cite{phenix-pp1,phenix-pp2} are self-explanatory, and the stars show the results of the parameter-free calculations with proper 
kinematics~\cite{kpss-spin}.
}
 \end{figure}

The parameter-free calculations agree with the PHENIX experiment 
data~\cite{phenix-pp1,phenix-pp2}, as  
demonstrated in Figure~\ref{fig:AN-pi-a}.

\section{Polarized \boldmath$pA$ Collisions}

\subsection{The Cross Section 
}
{\subsubsection{Glauber approximation}} 

A natural extension of Equation (\ref{100}) to nuclear targets has the form
\beqn
\frac{d\sigma(pA\to nX)}{d\ln(z)dq_T^2d^2b_A}&=&
f_{\pi/p}(z,q_T)
\nonumber\\ &\times&
\int  \frac{d\sigma^{\pi A}_{\rm{tot}}(M_X^2)}{d^2b_A}\,
 S_{NA}(b_A),
\label{580}
\eeqn
where $b_A$ is the impact parameter of $pA$ collision, and $S_{NA}(b_A)$ is the additional nuclear absorption factor described below.

This expression can be interpreted as an interaction of the projectile 
Fock component, $|\pi^+n\ra$,   
with the target, whereby the proton light-cone momentum in the respective fractions of $z$ and $(1-z)$ are 
shared. While the pion interacts inelastically with the target, the spectator neutron has to remain intact, i.e., it has to survive 
{its passage} 
through the nucleus.

The partial total pion-nucleus cross section in Equation~(\ref{580}) can be evaluated in the Glauber approximation~\cite{mine,kps}
\beqn
\left.\frac{\sigma^{\pi A}_{\rm{tot}}(M_X^2)}{d^2b_A}\right|_{\rm{Gl}} 
\approx
2\left[1-
e^{-{1\over2}\,\sigma^{\pi N}_{\rm{tot}}(M_X^2)\,
T_{\pi A}(b_A)}\right].
\label{620}
\eeqn

Here, 
 \beq
  T_{\pi A}(b_A)=\int d^2\Delta\,f_{\pi N}(\Delta)\,T_A(\vec b_A-\vec \Delta),
 \label{620a}
 \eeq
 where
$
T_A(b_A) = \int_{-\infty}^\infty d\zeta\,\rho_A(b_A,\zeta)$, is the nuclear thickness function, and $\rho_A(b_A,\zeta)$
 is the nuclear density that is used in calculations in the Woods-Saxon form. 
 The normalized partial pion-nucleon elastic amplitude is approximated by the
 Gaussian form
 \beq
 f_{\pi N}(\Delta)=\frac{1}{2\pi B^{\pi N}_{\rm{el}}}
 e^{-\Delta^2/2B^{\pi N}_{\rm{el}}},
 \label{620b}
 \eeq
 where the elastic slope rises with energy that is approximately expressed as $B^{\pi N}_{\rm{el}}(s_{\pi N})=
 B_0+2\alpha^\prime_{\Pom}\ln(s_{\pi N}/s_0)$ with $B_0=6\GeV^{-2}$, $\alpha^\prime_{\Pom}=0.25\GeV^{-2}$, $s_0=1\GeV^2$~\cite{kpss}.
 Notice that, for numerical calculations we use hereafter that are more 
 accurate than the optical approximation form, we replaced $e^{-{1\over2}\,\sigma^{\pi N}_{\rm{tot}} 
T_{\pi A}} \Rightarrow [1-\sigma^{\pi N}_{\rm{tot}} T_{\pi A}/2A]^A$, where the quantity $A$ denotes the atomic mass number.

Correspondingly, the neutron survival factor in (\ref{580}) reads as follows:
\beqn
S_{NA}(b_A)\bigr|_{\rm{Gl}} &\approx&
\frac{e^{-\sigma^{nN}_{\rm{in}}(zs)T_{NA}(b_A)}-e^{-\sigma^{pN}_{\rm{in}}(s)T_{NA}(b_A)}}
{T_{NA}(b_A)[\sigma^{pN}_{\rm{in}}(s)-\sigma^{nN}_{\rm{in}}(zs)]}
\nonumber\\ &\approx& 
e^{-\sigma^{NN}_{\rm{in}}(s)T_{NA}(b_A)}.
\label{800}
\eeqn

 If the nucleus remains intact or decays into fragments without particle production, 
then instead of the total $\pi A$ cross section, which one should use in (\ref{580}), the 
diffractive cross sections related to elastic $\pi A\to\pi A$ and quasielastic $\pi A\to\pi A^*$ channels should be used instead.
The corresponding cross section has the form ~\cite{mine,kps}
\beqn
\left.\frac{\sigma^{\pi A}_{\rm{diff}}(M_X^2)}{d^2b_A}\right|_{\rm{Gl}} &=&
\left[1-e^{-{1\over2}\sigma^{\pi N}_{\rm{tot}}T_A(b_A)}\right]^2
\nonumber\\ &+&
\sigma^{\pi N}_{\rm{el}}T_A(b_A)e^{-\sigma^{\pi N}_{\rm{in}}T_A(b_A)},
\label{890}
\eeqn
where the first and second terms correspond to the elastic and quasi-elastic scatterings,
respectively.

The difference between the cross sections, as in Equations~(\ref{580}) and (\ref{890}), is related to 
the inelastic 
 $\pi A$  
interactions, which lead to multiparticle production. Correspondingly, one should modify Equation (\ref{580}) by replacing the total by 
inelastic $\pi A$ cross section~\cite{mine,kps},
\beq
\left.\frac{\sigma^{\pi A}_{\rm{in}}(M_X^2)}{d^2b_A}\right|_{\rm{Gl}} =
1-e^{-\sigma^{\pi N}_{\rm{in}}T_A(b_A)}
\label{885}
\eeq

\subsubsection{Gribov corrections: color transparency}

It is known that the Glauber approximation is subject to Gribov inelastic shadowing corrections~\cite{gribov69},
which are known to make the nuclear matter more transparent for hadrons~\cite{zkl,mine}; in addition, these corrections also affect both factors in Equation (\ref{580}), as well as suppress $\sigma^{\pi A}_{\rm{tot}}$ and increase $S_{NA}(b_A)$.
We calculated the Gribov corrections to all of the orders of the multiple interactions by employing the dipole representation, as is described 
in Refs.~\cite{zkl,mine,kps,gribov85}. 

Hadron wave function on the light front can be expanded over different Fock states, and they consist of parton 
ensembles that have the various transverse positions $\vec r_i$ of the partons. The interaction cross section of such a 
hadron is averaged over the Fock states $\sigma_{\rm{tot}}^{hp}=\la\sigma(\vec r_i)\ra_h$, where $\sigma(\vec r_i)$ is the cross section of the interaction
of the partonic ensemble when it has the transverse coordinates $\vec r_i$ with the proton target.

Notice that high-energy partonic ensembles are eigenstates of interaction~\cite{zkl}, i.e., the parton coordinates $\vec r_i$ remain unchanged 
during the interaction. Therefore, the eikonal approximation employed in the Glauber 
 model
 (\ref{620})--(\ref{890}) 
should not be used for hadrons. However, with respect to the 
Fock components, the whole exponential terms
should be averaged~\cite{zkl,gribov85}. This corresponds to the following replacements in Equations~(\ref{620})--(\ref{890}): 
\beq
e^{-{1\over2}\sigma^{h N}_{\rm{tot}}T_A}=
e^{-{1\over2}\left\la\sigma(\vec r_i)\right\ra_h\,T_A}
\Rightarrow
\left\la e^{-{1\over2}\sigma(\vec r_i)\,T_A}\right\ra_{\rm h} \,, 
\label{895}
\eeq
where subscript 'h' indicates averaging over hadrons.

The difference between these averaging procedures can be exactly represented by the Gribov corrections~\cite{zkl,gribov85}.

The results of averaging in Equation (\ref{895}) for the proton-nucleus interactions was calculated in 
Ref.~\cite{kps} with a realistic saturated parameterization of the dipole cross 
section~\cite{kst2}; in addition, the quark-diquark model was used for the nucleon wave function: 
\beqn
\left\la e^{-{1\over2}
\sigma(r_T)T_A}\right\ra
=e^{-\frac{1}{2}
\sigma_0\,T_A(b)}
\sum\limits_{n=0}^\infty
\frac{[\sigma_0\,T_A(b)]^n}
{2^n\,(1+n\,\delta)\,n!},
\label{370}
 \eeqn
where 
\beq
\sigma_0(s)=\sigma_{\rm{tot}}^{pp}(s)\left[1+
\frac{1}{\delta}\right],
\label{375}
\eeq
and $\delta= 8\la r_{\rm{ch}}^2\ra_p/3R_0^2(s)$.
{We} {use} 
{the mean proton charge radius squared} 
$\la r_{\rm{ch}}^2\ra_p=0.8\fm^2$~\cite{r-ch} and the energy-dependent saturation radius 
 $R_0(s)=0.88\,{\rm fm}\,(s_0/s)^{0.14}$ with $s_0=1000\,\text{GeV}^2$ ~\cite{kst2}.  

Gluon shadowing corrections, which correspond to the triple-Pomeron term in the context of diffraction, were 
also introduced~\cite{mine,kps}.

\subsection{{Calculation Results}} 

\begin{figure}[H]
    \scalebox{0.3}{\includegraphics{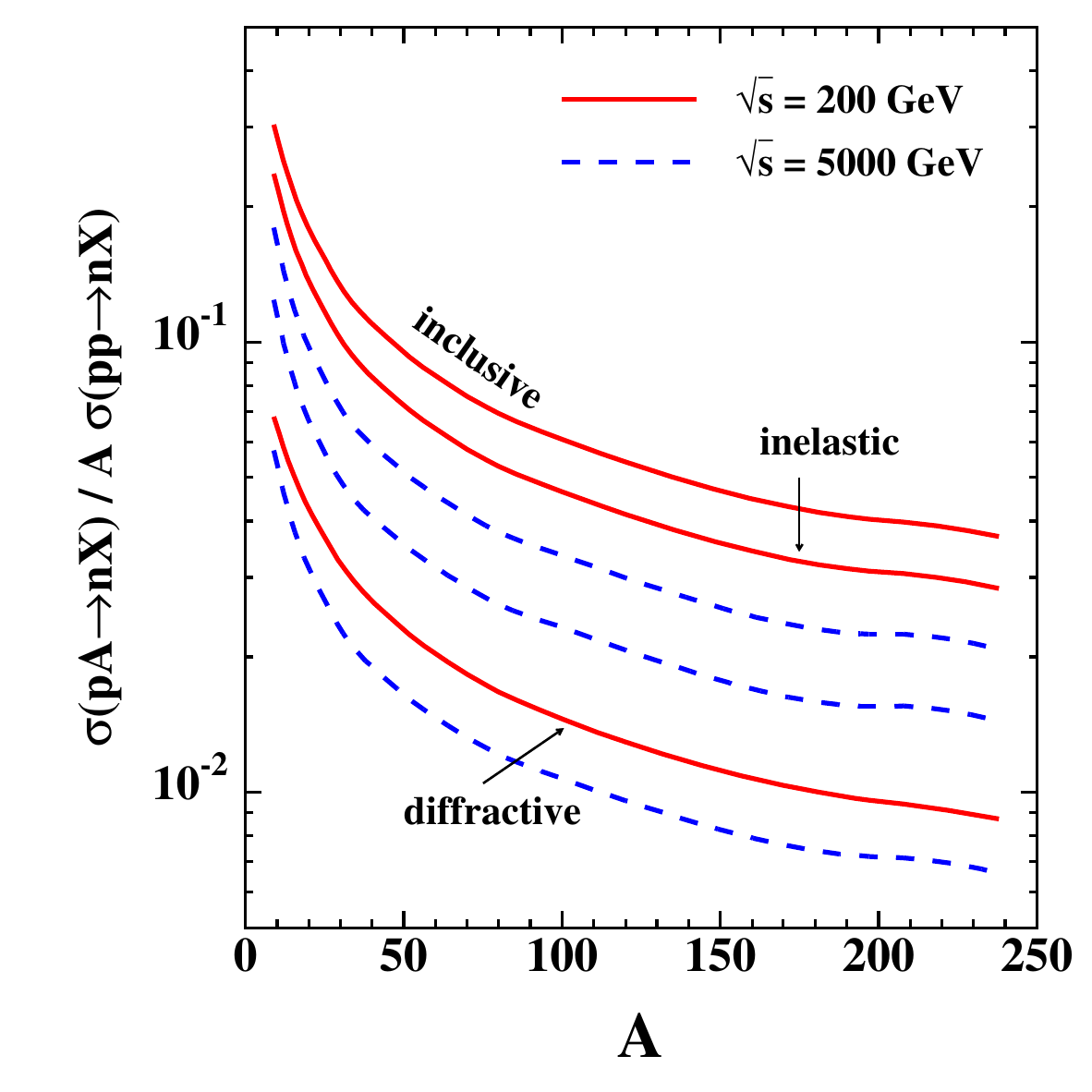}}
\caption{\label{fig:sig2} The $pA$ impact parameter $b_A${-integrated} 
{cross} sections, which are normalized by the $pp$ cross section of the leading neutron production. 
The solid and dashed curves correspond to both the $pAu$ and $pPb$ collisions at $\sqrt{s}=200 \GeV$ and $5000 \GeV$, respectively. 
The three curves in each set 
 correspond to the inclusive, inelastic 
and diffractive neutron productions (top to bottom).
 }
 \end{figure}

\begin{figure}[H]
  \scalebox{0.3}{\includegraphics{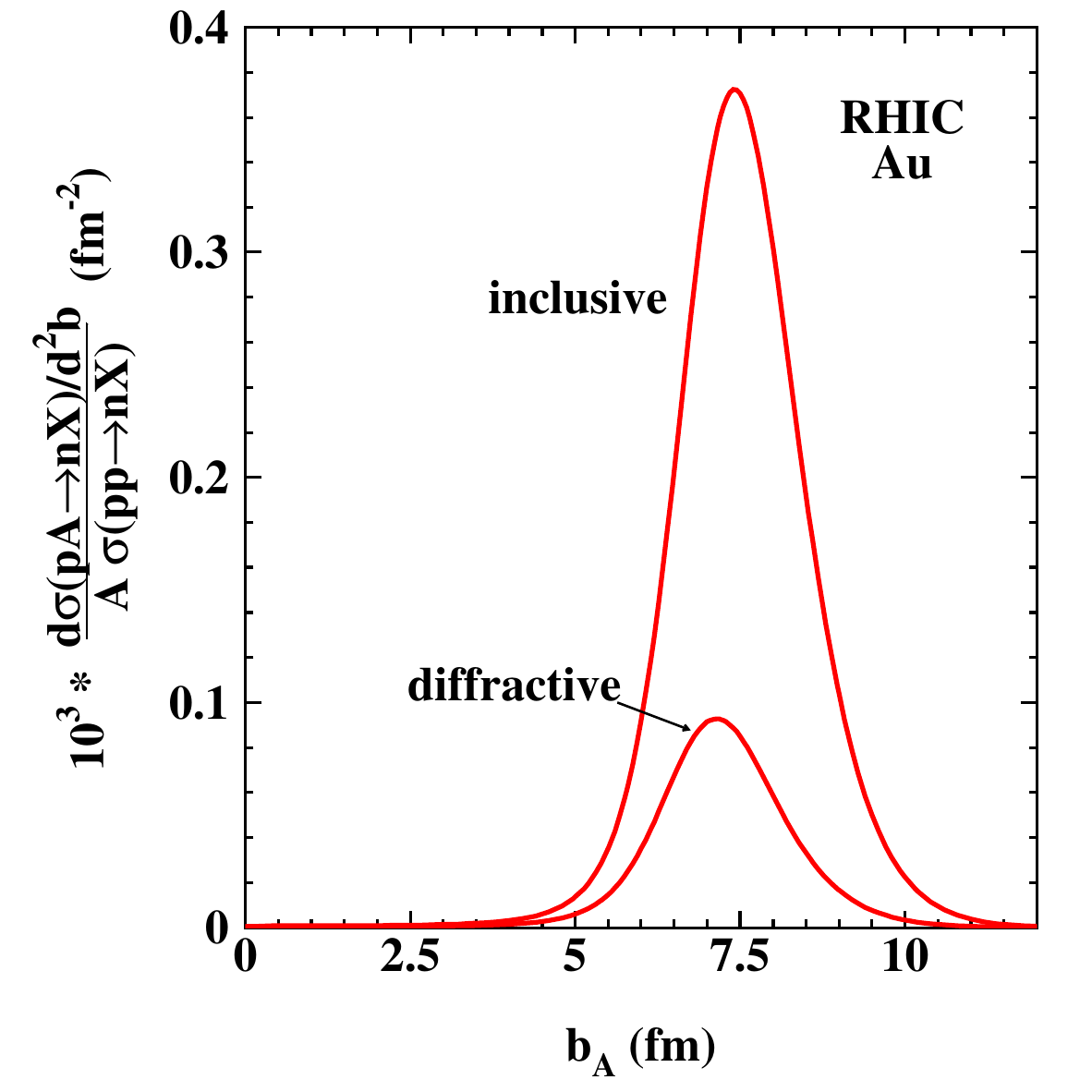}}
\caption{\label{fig:sig1} Partial cross sections for the inclusive 
and diffractive neutron
 production in the $pAu$ collisions at $\sqrt{s}=200 \GeV$ and $z=0.75$.}
 \end{figure}

The results of the Glauber model calculations, including the Gribov corrections, are obtained for the partial inclusive cross section of $pAu\to nX$. Furthermore, Equation~(\ref{580}) is normalized by the $pp\to nX$ cross section, and this is shown in Figure~\ref{fig:sig2} 
 via the top solid red (RHIC) and top blue dashed (LHC)~curves.
The two other lower curves show the cross section of the inelastic and diffractive channels.
One can see that the cross section is quite small, and this can be understood as a consequence of
the significant suppression that occurred by a factor of $S_{NA}(b_A)$, as can been seen in Equation~(\ref{800}). Indeed, the impact parameter dependencies of the 
inclusive (\ref{620}) and diffractive (\ref{890}) cross sections of neutron production are depicted in Figure~\ref{fig:sig1}. One can see that neutrons are produced from the very periphery of the nucleus, and this is why the $b_A$-integrated cross section is so small. Correspondingly, the ratio shown in Figure~\ref{fig:sig2} fell for the heavy nuclei as $A^{-2/3}$.

When close to the pion pole, we treated the vertex $\pi+p\to X$, as shown in Figure~\ref{fig:3R}, as the pion-proton interaction amplitude.
By squaring it, one obtains the total pion-proton cross section, as was obtained in 
Equation (\ref{100}). In the case of a nuclear target, one should replace the proton by a nucleus and obtain $\sigma_{\rm{tot}}^{\pi A}(M_X^2)$.
The latter obtained 
contributions from different channels, which can be classified as inelastic and diffractive interactions.
The former corresponds to multiparticle production, which filled the rapidity interval between the colliding pion and nucleus, while the latter corresponds to the rapidity gap events.
The corresponding cross sections can be evaluated with the models described below.

The recent measurements of the forward neutrons in the PHENIX {experiment}
~\cite{prl,bazil1,itaru} was supplemented with beam-beam counters (BBCs), whereby the charged particles in the two pseudo-rapidity intervals of $3.0<|\eta|<3.9$ were detected. The results of the measurements of the forward neutrons are
presented here for the different samples of events:
\begin{itemize}
\item[(i)]	
inclusive neutron production with the BBCs switched off; 
\item[(ii)]	
neutrons accompanied with multiparticle production (where one or both BBCs are fired), can be associated with inelastic
$\pi A$  
collisions;
\item[(iii)]	
if
both BBCs are vetoed, a large contribution of diffractive interactions might be expected.
\end{itemize}


Notice that such a correlation with BBC activities and related processes should not be taken literally, and that a comparison with theoretical predictions should be performed with precaution. Further experimental studies employing Monte-Carlo simulations are required.

\section{Glauber Model for Single-Spin Asymmetry in Polarized \boldmath$pA\to nX$}

Single-spin asymmetry that occurs on a nuclear target due to 
 $\pi\, \tilde a_1$  
interference can be calculated with a modified Equation (\ref{920}), in which one should replace
\beq
\frac{d\sigma_{\pi p\to \tilde a_1p}(M_X^2)/dt|_{t=0}}
{d\sigma_{\pi p\to\pi p}(M_X^2)/dt|_{t=0}}
\Rightarrow
\frac{d\sigma_{\pi A\to \tilde a_1A}(M_X^2)/dt|_{t=0}}
{d\sigma_{\pi A\to\pi A}(M_X^2)/dt|_{t=0}} \, .
\label{500}
\eeq

The replacement (\ref{500}) 
leads to a single-spin asymmetry, which can be presented in the~form
\beq
A_N^{pA\to nX} =
A_N^{pp\to nX}\times \frac{R_1}{R_2}\,R_3.
\label{520}
\eeq 

In the ratio $R_1/R_2$, 
 {$R_1$}  
 corresponds to the numerator and $R_2$ to the denominator in the right-hand side  
of Equation 
 {(\ref{500})}. 
The factor $R_3$ is related to the experimental setup and is fixed below.

Factor $R_1$, in accordance with Equations 
(\ref{920}) and (\ref{500}), is the nuclear modification factor for the forward amplitude of $\pi A\to \tilde a_1 A$, which is a coherent diffractive transition. In the Glauber approximation, it has the form
\beqn
&&R_1=\int d^2b_A\int\limits_{-\infty}^\infty d\zeta\,
\rho_A(b_A,\zeta)\, e^{-{1\over2}\sigma^{pp}_{\rm{tot}}T_A(b_A)}
\nonumber\\ &\times&
\exp\left[-{1\over2}\sigma^{\pi N}_{\rm{tot}} T_-(b_A,\zeta)
-{1\over2}\sigma^{\tilde a_1 N}_{\rm{tot}} T_+(b_A,\zeta)\right]\!,
\label{525}
\eeqn
where $T_-(b_A,\zeta)=\int_{-\infty}^\zeta d\zeta^\prime\, \rho_A(b_A,\zeta^\prime)$ and $T_+(b_A,\zeta)=
T_A(b_A)-T_-(b_A,\zeta)$.

In Equation
(\ref{525}), we assume that the incoming pion propagates through the nuclear thickness 
$T_-(b_A,\zeta)$, and that the outgoing $\tilde a_1$ propagates through $T_+(b_A,\zeta)$. Both attenuate with their corresponding cross sections. The transition occurs at the point with the longitudinal coordinate $\zeta$, which varies from $-\infty$ to $+\infty$.

When integrating Equation (\ref{525})
over $\zeta$ analytically, 
one arrives at
\beqn
R_1&=& \frac{1}{\Delta\sigma}\int d^2b_A\, e^{-{1\over2}\sigma^{\pi p}_{\rm{tot}}T_A(b_A)}
 \nonumber\\ &\times&
\left[1-e^{-{1\over2}\Delta\sigma T_A(b_A)}\right]
 e^{-{1\over2}\sigma^{pp}_{\rm{tot}}T_A(b_A)},
 \label{540}
 \eeqn
 where $\Delta\sigma=\sigma^{\tilde a_1 N}_{\rm{tot}}-\sigma^{\pi N}_{\rm{tot}}$.
 As was mentioned above, and was explained in detail in
Refs.~\cite{kpss-spin,belkov,pcac}, the diffractive
production of the $a_1$ axial-vector meson is a very weak signal compared with $\rho$-$\pi$ production, which form a rather narrow peak in the invariant mass distribution.
Therefore, $\sigma^{\tilde a_1 N}_{\rm{tot}}=\sigma^{\rho N}_{\rm{tot}}+\sigma^{\pi N}_{\rm{tot}}$ and
$\Delta\sigma=\sigma^{\rho   N}_{\rm{tot}}$ have 
{acceptable} 
accuracy. The data for the photoproduction of the $\rho$ meson in the nuclei agree with
$\sigma^{\rho   N}_{\rm{tot}}\approx \sigma^{\pi   N}_{\rm{tot}}$; as such, we fix $\Delta\sigma$ at this value.

 Nuclear modification, which corresponds to the denominator of Equation~(\ref{500}), is determined by
Equations~(\ref{580})--(\ref{800}), and it has the form
 \beqn
 R_{2}= \frac{2}{\sigma^{\pi p}_{\rm{tot}}}\int d^2b_A
\left[1-e^{-{1\over2}\sigma^{\pi p}_{\rm{tot}}T_A(b_A)}\right]
 e^{-{1\over2}\sigma^{pp}_{\rm{tot}}T_A(b_A)}.
 \label{560}
 \eeqn
  
\begin{figure}[H]
 {\includegraphics[height=6cm]{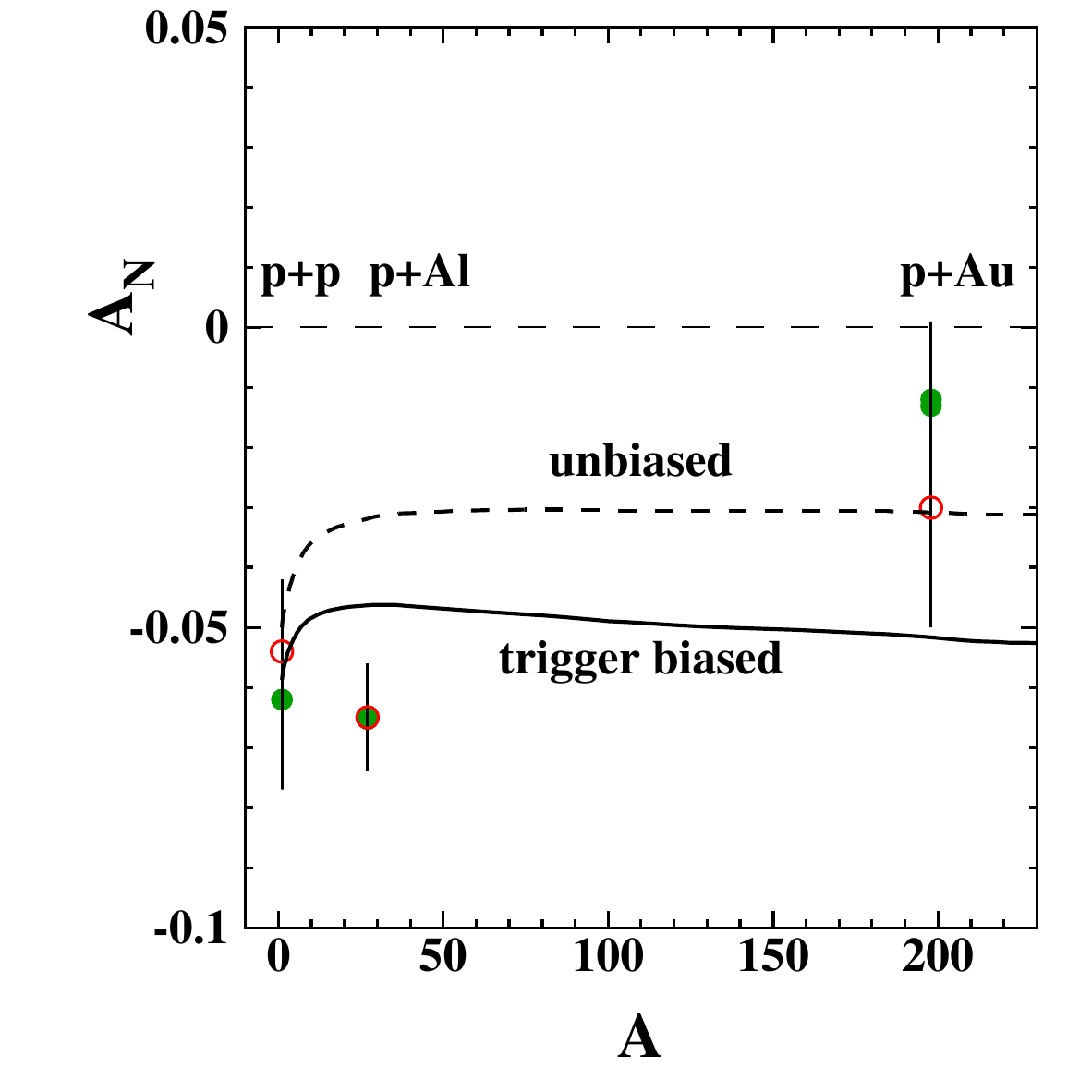}}
\caption{\label{fig:AN} $A_N$ {in polarized} 
 $pA\to nX$ 
 versus 
$A$ at $\sqrt{s}=200 \GeV$, $\la q_T\ra=0.115 \GeV$, and $\la z\ra = 0.75$. The full and open data points correspond to events with either both 
BBCs fired or with only one of them fired in the nuclear direction, respectively~\cite{prl,bazil1,itaru}. An attempt to model these two classes 
of events is represented by the solid and dashed curves, respectively. 
See text for details.
}
 \end{figure}

The factor $R_3$ depends on how the measurements were conducted.
If the BBCs are fired, a proper estimate would be 
$R_3=\sigma^{\pi A}_{\rm{tot}}/\sigma^{\pi A}_{\rm{in}}$. 
Otherwise, if the BBC are switched off (i.e., inclusive neutron productions), we fix $R_3=1$. The results corresponding to these two choices are 
plotted in Figure~\ref{fig:AN} by solid and dotted curves, respectively. 
All the data points that correspond to the events with BBCs were fired. However, the full green and open red points corresponded to events where either both BBCs fired or only one of them fired in the nuclear direction, respectively~\cite{prl,bazil1,itaru}.


The difference between these two results reflects the uncertainty in the physical interpretation of events when there were fired with vetoed 
BBCs. This can be improved by applying a detailed Monte Carlo modeling. Nevertheless, the results of our calculations,
as presented in Figure~\ref{fig:AN}, can be reproduced reasonably well in experimental data~\cite{prl,bazil1,itaru}.

\begin{figure}[H]
  \scalebox{0.3}{\includegraphics{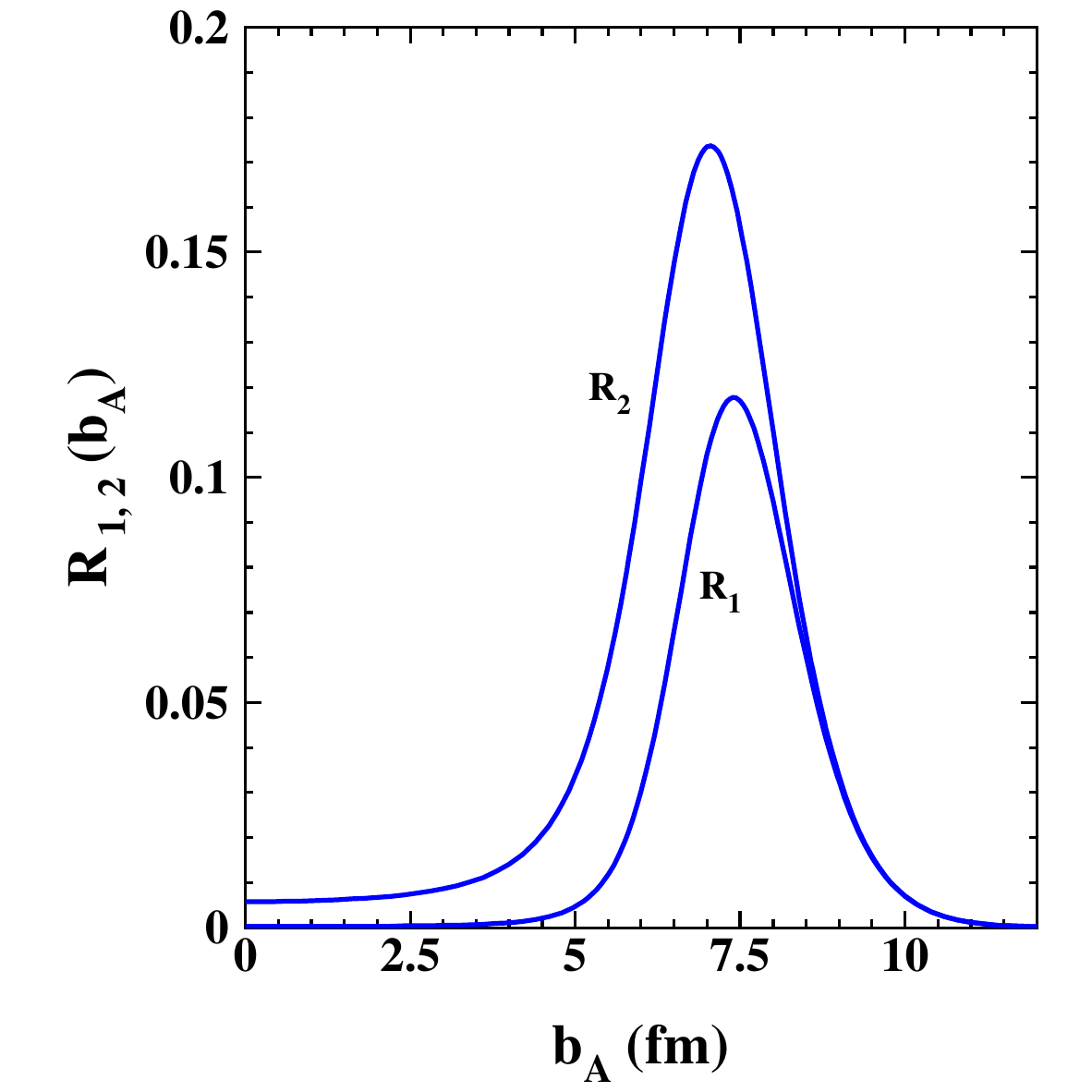}}
\caption{\label{fig:R12} Impact parameter dependence of the denominator $R_1$
and numerator $R_2$ in Equation (\ref{520}) according to Equations  
(\ref{540}) and (\ref{560}), respectively. Calculations are performed for 
neutron production in the $pAu$ collisions at $\sqrt{s}=200\GeV$.}
 \end{figure}

A remarkable feature of the single-spin asymmetry $A_N$ of the neutrons produced on nuclear targets is 
quite a weak $A$-dependence, which is seen both in the data and in our calculations. The reason for this can be 
understood as follows. All of the $A$-dependence of the $A_N$ asymmetry is contained in the factors $R_1$ and $R_2$ in Equation~(\ref{520}). It turns out that the strong nuclear absorption factor $S_{NA}(b_A)$ in Equation~(\ref{800}) is contained in both Equations (\ref{540}) and (\ref{560}), and this factor pushes
neutron production to the very periphery of the nucleus. This is demonstrated by the $b_A$-unintegrated factors $R_1(b_A)$ and $R_2(b_A)$, which are plotted in Figure~\ref{fig:R12} in gold at $\sqrt{s}=200\GeV$. Due to the observed similarity of the $A$-dependencies of $R_1$,  $R_2$, and 
 {proportional to} 
$
A^{1/3}$, they are mostly canceled in Equation 
(\ref{520}), thereby resulting in a nearly $A$-independent single-spin asymmetry of neutrons.

\vspace{-6pt}

\section{Summary}

The previously developed methods of calculation of the cross section of the leading neutron production in $pp$ collisions were extended to nuclear targets. The nuclear absorptive corrections, as calculated in the Glauber-Gribov approach, were so strong that they pushed the partial cross sections of the leading neutron production to the very periphery of the nucleus. As a result, the $A$-dependencies of the inclusive and diffractive neutron production turned out to be similar, i.e., $\sim$$A^{1/3}$. 

The mechanism of the 
$\pi\, a_1$ 
interference, which successfully explains the observed single-spin asymmetry in the polarized reaction $pp\to nX$, was extended to the  collisions of the polarized protons with nuclei. When corrected for nuclear effects, it 
explained 
well {enough} 
the observed $A_N$ asymmetry in inelastic events, which is when the nucleus violently breaks up~\cite{prl,bazil1,itaru}. However, the large value and the opposite sign of the $A_N$ observed in the diffractive sample still remains a~challenge.

\vspace{--6pt}
\authorcontributions{{The authors contributed equally to this work.}}

\funding{{This} 
work was supported, in part, by the grants from Chilean National Agency for Research and Development {(ANID)}
---Chile {FONDECYT} 
1231062 and 1230391, by ANID {PIA/APOYO AFB}220004, and by ANID---The Millennium Science Initiative Program 
{ICN}2019\_044.}


\dataavailability{{All data used in this analysis have been published in 
the cited publications.}}

\acknowledgments{We are thankful to Alexander Bazilevsky and Itaru Nakagawa  
for providing us with data and the details of the measurements, as well as
for numerous informative {discussions.}
}

\conflictsofinterest{{The authors declare no conflict of interest.}} 


\begin{adjustwidth}{-\extralength}{0cm}
\reftitle{References}

\PublishersNote{}
\end{adjustwidth}

\end{document}